\documentclass[manuscript,screen]{acmart}

\usepackage{caption}
\usepackage{subcaption}

\AtBeginDocument{%
 \providecommand\BibTeX{{%
 \normalfont B\kern-0.5em{\scshape i\kern-0.25em b}\kern-0.8em\TeX}}}

\copyrightyear{2021}
\acmYear{2021}
\setcopyright{acmcopyright}\acmConference[Asian CHI Symposium 2021 ]{Asian CHI
Symposium 2021 }{May 8--13, 2021}{Yokohama, Japan}
\acmBooktitle{Asian CHI Symposium 2021 (Asian CHI Symposium 2021 ), May 8--13,
2021, Yokohama, Japan}
\acmPrice{15.00}
\acmDOI{10.1145/3429360.3468176}
\acmISBN{978-1-4503-8203-8/21/05}

\usepackage{csquotes}

\begin{document}

\title{Together we learn better: leveraging communities of practice for MOOC learners}


\author{Dilrukshi Gamage}
\affiliation{  
\institution{ Department of Computer Science and Engineering, University of Moratuwa}
\city{Moratuwa}
\state{}
\country{Sri  Lanka}}
\email{dilrukshi.gamage@acm.org}


\author{Mark E Whiting}
\affiliation{%
\institution{ The Wharton School, University of Pennsylvania}
\city{Philadelphia}
\state{Pennsylvania}
\country{USA}}
\email{markew@seas.upenn.edu}








\begin{abstract}
 MOOC participants often feel isolated and disconnected from their peers. 
 Navigating meaningful peer interactions, generating a sense of belonging, and achieving social presence are all major challenges for MOOC platforms. 
 MOOC users often rely on external social platforms for such connection and peer interaction, however, off-platform networking often distracts participants from their learning. 
 With the intention of resolving this issue, we introduce PeerCollab, a web-based platform that provides affordances to create communities and support meaningful peer interactions, building close-knit groups of learners. 
 We present an initial evaluation through a field study (n=56) over 6 weeks and a controlled experiment (n=22). 
 The result indicates insights on how learners build a sense of belonging and develop peer interactions leading to close-knit learning circles. 
 We find that PeerCollab can provide more meaningful interactions and create a community to bring a culture of social learning to decentralized, and isolated MOOC learners. 
\end{abstract}

\ccsdesc[500]{Human-centered computing~Collaborative and social computing systems and tools}

\keywords{MOOCs,Community of practice, Social Learning, Sense of belonging}

\maketitle

\section{Introduction}
Massive Open Online Courses (MOOCs) are revolutionizing the online learning landscape. MOOCs are different from traditional structured learning due to their open enrollment design and no strong obligation to complete courses. As a consequence, MOOCs see massive enrollment but often have very limited course completion rates --- nominally 10--20\%~\cite{reich2019mooc}. Although completing a course may not be every student's goal, some participants in MOOCs drop out due to a mismatch in interests, learning styles or undesirable learning experiences~\cite{wang2019effects}, while many more leave due to a lack of social support and attachment within the class~\cite{rose2014social}.

In any context (online or in person), participants learn effectively when they can interact with each other while socially engaging as a collective~\cite{wang2007effects}. Especially in online courses, fostering a strong sense of community among learners has been the goal of many instructors and is considered essential for supporting students' learning experiences~\cite{phirangee2017othering}. Many MOOC platforms hinder social learning due to their lack of design affordances that target this issue~\cite{gamage2020moocs}. Forum spaces in MOOC platforms are typically the only option for MOOC participants to interact with their peers. Yet, MOOC forums are themselves isolating, due to a massive number of conversations --- participants often find it difficult to navigate meaningful peer interactions~\cite{coetzee2014should}. Further, forum spaces in MOOCs are generally used only to ask questions or report errors to the teaching team. Students who expect much learner engagement and interaction in a MOOC have no option, and rely on friend groups or off-platform social networking to cultivate connections for learning and create social capital~\cite{zheng2016role}. 

Groups features on Facebook, LinkedIn, or WatsApp, as well as Twitter lists are common social spaces for MOOC participants to build learning communities. Occasionally, learners also meet face-to-face learning circles in P2P~\cite{sun2019distance}. However, these communities all face challenges with constant distractions entailed through other activities which large scale social platforms are optimized to support, and with physical meetups, groups face challenges around scale, proximity and scheduling. Although online communities for learning are extremely important, successful online communities must overcome challenges common to off-line groups: e.g., recruiting and socializing new members, developing members’ commitment, eliciting contributions, regulating behavior and coordinating work. Thus these communities are highly dependent on how they are designed --- how participants navigate interests, motivation, and sense of community for an effective learning experience. Although there are multiple successful online communities for learners, embedded social technology designs for MOOCs remain unresolved by human-computer interaction research and industry. 

In this paper, we introduce PeerCollab, a system for MOOC learners to find a group of learners interested in studying together while developing social capital --- creating a sense of belonging and enhancing peer interaction and learning experience. Design inspiration to build PeerCollab was drawn from the theories of Communities of Practice (CoP) introduced by social theorist Wenger~\cite{wenger2011communities} who describes a learning theory with a strong relationship to the social construction of knowledge. Early stages of the platform were evaluated using a control study to understand if the platform enables sense of belonging and a field study to understand how groups of learner communities learn using PeerCollab.

We present results indicating a significant level of progress in building a sense of belonging among learners. We also surface insights about how peer interactions are shaped in a structured community using socio-technological systems. In general, MOOC platforms isolate decentralized learners limiting then from interacting meaningfully and prohibiting social learning. PeerCollab introduces a change to the didactic learning culture by scaffolding a social learning culture with structured online learning programs. It is optimized to handle the scale of MOOCs and we envision participants in PeerCollab making connections leading to lifelong networks supporting each other.

\section{Related Works}
\subsection{Socio-technical Systems to bring a sense of community and social capital to MOOCs}
The design of effective social interaction features for MOOCs is an ongoing research endeavour. For example, Talkabout allowed diverse groups of learners to interact socially over a video conferencing~\cite{kulkarni2015talkabout}, while other tools improve synchronous group interaction on MOOCs through gamification~\cite{antonaci2019gamification}, teamwork~\cite{wen2017supporting} and structured peer learning~\cite{coetzee2015structuring}. However, none of these has been broadly adopted by MOOC platforms because of the underlying technical or logistical resources required to use them at the scale of a large MOOC with thousands of synchronous learners. Asynchronous interaction scaffolds have also been explored. \citeauthor{kizilcec2014encouraging} encouraged forum participation with behavioral prompts, e.g., motivating collectivist approach (“your participation benefits everyone”), individualist (“you benefit from participating”), or neutral (“there is a forum”)~\cite{kizilcec2014encouraging}. Yet such encouragements are challenged by the pseudonymous nature of MOOC participants --- there remains limited sense of community in online courses and due to their general emphasis on individual achievement and limited duration~\cite{kizilcec2014encouraging}. Researchers claim that, when effective, forums enhance student learning, sense of community, and collaborative dialogues~\cite{luca2004using}. Similarly, research to understand facilitation through small groups~\cite{lim2014initial}, empirical research work on supportive technologies for group discussion in MOOCs provide substantial evidence of how discussions could redirect to effective learning in collaborative learning environments~\cite{rose2015supportive}. In particular, the case study of~\citeauthor{rose2015supportive} revealed insights into the limitations of learning through discussion in current MOOCs, as well as the instructors challenges to overcome those limitations without community support~\cite{rose2015supportive}. 

Community discourse research offers convincing evidence on the value of discussion for learning and social engagement~\cite{mitchell2011profound, resnick2015socializing}, yet in MOOCs, participants seldom post or reply and identify as "lurkers"~\cite{ferguson2015examining, clow2013moocs} --- very few learners engage in regular forum use. As a consequence, more empirical understanding of structuring and cultivating social behavior on MOOC platforms is needed. A particularly valuable target is leveraging community development methods to scaffold social learning for better learning experience overall within MOOCs. In their current state, most MOOC social features are provided through online forums with threaded discussion of posts and replies. Forum use is imperative to achieve effective learning experience, but investigations of MOOC forums show struggles to retain users over time~\cite{coetzee2014should} --- half of those initially active on these forums quickly become dormant, due to a lack of facilitation, overwhelming numbers of messages, and behavioral differences between more and less active students~\cite{mak2010blogs}. Since forums are disorganized, and hard to navigate, closer interactions leading to social engagement rarely develop~\cite{almatrafi2018systematic}, instead, informal groups associated with MOOC courses tend to emerge off-platform, on social media sites, as a way for learners to build their own social learning communities~\cite{koutsakas2018exploring}. Although social media provides a rich engagement for participants~\cite{zheng2016role}, these communities end up serving as more of a distraction distracting from the learning materials, and serve as a secondary isolation mechanism for students unaware of their existence~\cite{salmon2015space}. 

\subsection{Designing Communities for MOOCs}
Although numerous MOOOC communities exist on social networking sites across the internet such as Facebook, Twitter, reddit, these platforms are designed to engage users in other kinds of pursuit as well, leading to quick distraction from learning goals~\cite{manca2013tool}. MOOCs communities in a physical space have also been explored by analysing 4,000 MOOC-related events on the meetup.com website, a networking platform for coordinating in-person meetings~\cite{bulger2015real}. This study concluded that learners around the world engaged with co-located communities primarily to discuss content-related matters~\cite{bulger2015real} and another similar in person MOOC community revealed cognitive, social, and accountability gains for participants, but faced logistical challenges for scheduling and finding physical space~\cite{chen2015mooc,damasceno2018new}. Socio-technical systems that enrich the social experience of MOOC participants are, so far, very limited. PeerCollab is engineered considering theories underpinning successful community designs, specially we drew inspiration from Communities of Practice (CoP)~\cite{wenger2010communities} literature to engage social presence and sense of community. This theory explains what is required for knowledge building communities that keep human identity and social structure and their core. According to \citeauthor{wenger2010communities}, humans are social creatures, and learn more effectively when they take learning as social participation~\cite{gauvain2007socialization}. CoP can be use as a tool to build social learning communities and \citeauthor{wenger2010communities} describes 3 required constructs: 1) Domain, 2) Community and 3) Practice~\cite{wenger2010communities}. We designed PeerCollab to leveraging these constructs to build socio-technical system that bring social interactions for MOOC participants.

\begin{figure}[tb]
 \centering
 \includegraphics[width=\linewidth]{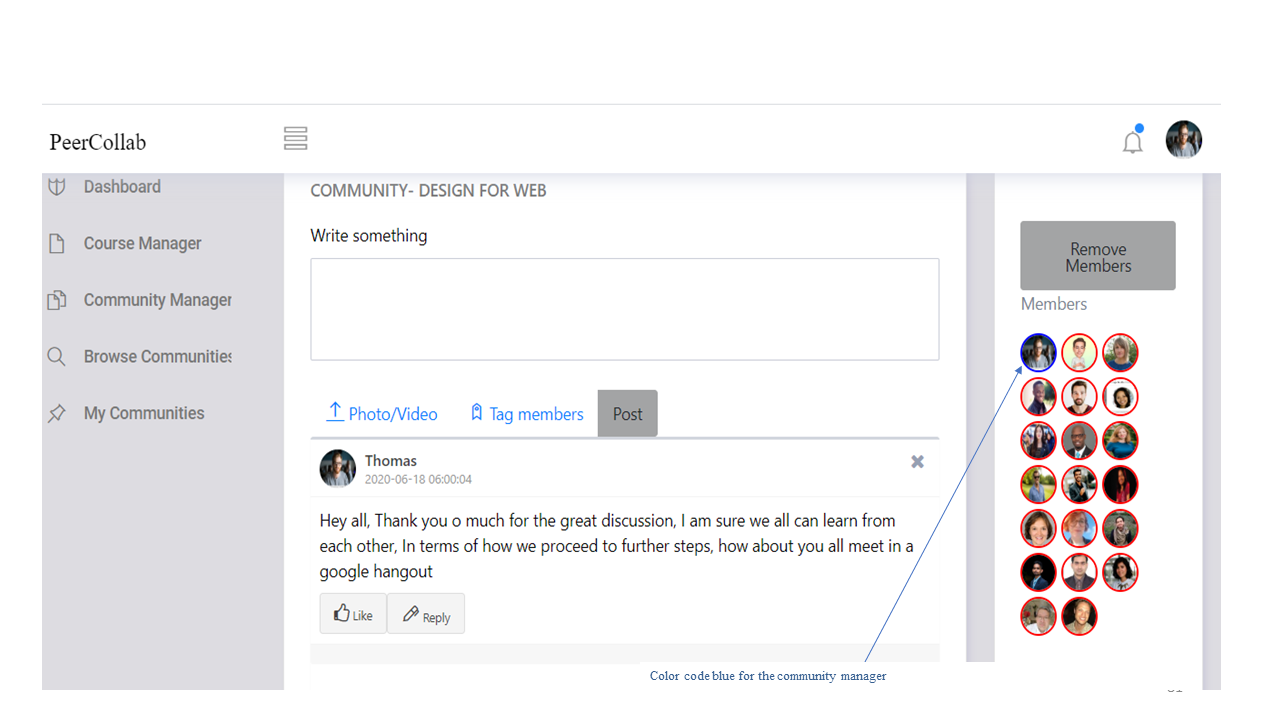}
 \caption{The interface inside a rapid community space in the PeerCollab system. Members are highlighted in pink and the community manager is highlighted in blue. Each member has a profile accessible to members of the community which helps to network with each other with personal message}
 \label{fig:join}
\end{figure}

\section{PeerCollab}
Inspired by social learning knowledge construction from the Communities of Practice (CoP) literature~\cite{wenger2010communities}, PeerCollab builds on core concepts of "Domain", "Practice" and "Community" to apply to the context of online learning. Our system provides 1) a domain for a group of online learners to create a community, 2) a practice for facilitation, and system features supporting cultivation of community practices and 3) a community to learn together throughout the MOOC course they decide to take as a group. The PeerCollab system is designed with features that facilitate learners to a) create a community and lead a group of learners, b) find learner communities with shared learning goals, c) join communities, and d) engage in leader-driven community activities. 

Together, these capabilities enable communities of practice where MOOC learners are provided opportunities to build social capital and learn together, enhancing peer interaction, social engagement and a sense of community. The system was built using the React framework for the front-end and Node.js for the back-end, a mySQL database. The following engineered features in the PeerCollab collectively enable MOOC participants to contribute to a social process leading to a sense of community.

\subsection{Finding communities with shared learning goals}
The PeerCollab platform allows any interested participant to join as a member of the platform and search for study groups we call "rapid communities". Rapid communities are listed with their description and size and enable community engagement in a number of threads under any category. 

\subsection{Joining a community of learners}
Members of the platform can join any rapid community. Rapid community members are asked to adhere to the community guidelines and a community leader (generally the person who started a given community) will facilitate awareness around these issues. Joining a rapid community will lead members to a unique space in the platform with features to communicate with that group. Forum spaces, notice boards, and other tools for community facilitation are all available. Each member of the community is visible and distinguished by a color code from the community manager (Fig.~\ref{fig:join}). 

\subsection{Lead a community of learners}
Any member of the PeerCollab can create a rapid community with the intention of leading a group of interested learners who may join as community members. Rapid communities require a unique name, description, and information on the study group and its associated learning goals. The creator of the community is identified as the community manager and receives instructions on community manager duties and responsibilities.

\section{Evaluation}
We evaluate PeerCollab with two key questions: 1) Does PeerCollab increase the sense of belonging and social presence in a MOOC more than a standard MOOC forum? 2) What are the behaviors reflected in PeerCollab rapid communities? We conduct a controlled study to explore the first question and a field study to explore the second. 

\subsection{Controlled study}
We conducted a randomized between-subject experiment, where a control group takes a course in a MOOC platform using the built-in forum space, and a treatment group takes the same course on the same platform, with PeerCollab enabled. We recruited 22 participants through social sampling, targeting individuals familiar with MOOCs and having taken at least one course on Coursera, edX or OpenHPI MOOC platforms. The control condition had 5 females, 6 males, while the PeerCollab condition had 7 males and 4 females. Participants' age ranged from 21-40. 

Before the experiment, participants ranked 3 courses options: \textit{Creative problem solving} on Coursera, \textit{Human Centered Design thinking} on OpenHPI and \textit{Designing User experience} on edX --- all 4 week courses that were scheduled to start shortly after the recruitment period and that were popular and had no prerequisite requirements. 75\% of our participants preferred \textit{Creative problem solving}, so this was chosen as the focal MOOC for our study. 

Control participants used all standard Coursera features, and were instructed to share their forum activity during the experiment, e.g., all peer interactions and postings. PeerCollab participants joined the same course on Coursera, but were instructed to use PeerCollab for their social participation. When joining PeerCollab, they were added to an initial rapid community which provided them with a brief introduction to the platform features. Some members self selected to operate as community managers and received instruction about the platform and facilitation specific to that role.

At the end of the 4 week course, each participant in both conditions completed a survey measuring sense of belonging based on a validated scale~\cite{ribera2017sense,sense@site} with adaptations for PeerCollab and reducing from 10 to 6 items based on platform relevance. Social presence was evaluated by analyzing the content of the interactions during the course. Interactions were coded as forum messages, likes to the posts in the forums and online meeting conversations. Control participant data were obtained as self-reports from members and the PeerCollab participant data were obtained from our platform. The experiment was under our institutes' ethical clearance and we handled data anonymously by giving a unique identifier to each user. 

\subsection{Field study}
To explore how would participants create and join communities and adhere to the structure we created, we examined their behaviors to understand if it reflects the qualities of communities of practice intended by the design of PeerCollab. We recruited participants from MOOC communities hosted in Facebook groups and invited participants to sign up to trial PeerCollab. 
 For the purposes of a field analysis, we created an instance of PeerCollab offering support for \textit{Introduction to Java} on OpenSAP, \textit{Design thinking} on OpenHPI, \textit{Introduction to C++} on edX, and \textit{Creative Problem-solving} on Coursera --- all courses spanning 4--6 weeks in their respective platforms. 

56 participants joined PeerCollab for this evaluation over a period of one week. Participants consisted of 26 females and 30 males aging 23--45, with 68\% identifying as university or college students, freelancing/ working or occupied in a job. 32 joined a space for \textit{Design thinking}, 30 for \textit{Creative problem-solving}, two for \textit{C++} and 8 for \textit{Introduction to Java Course}. Members could join any space and were not restricted to joining only one space. Members created two communities for \textit{Design thinking} and one community in each other course. In the \textit{Design thinning}, one community had 26 members and other community had 6 members. After 6 weeks, we analyze the interactions that had occurred on PeerCollab. There were forum posts, likes, scheduled meetings, and exchanged artifacts within each community. We also asked participants to complete a post-survey to understand their sense of belonging. We randomly selected 5 participants from different communities for interviews on their experience. 



\section{Results}
\subsection{Does PeerCollab increase sense of belonging and social presence?}
Through our controlled experiment, we compared self-reported control participation and PeerCollab rapid community participation. Only 2 participants in the control condition reported posting a comment in the forum, while the PeerCollab condition had 92 threads in the community space. The challenges of self-report notwithstanding, this difference is stark, and reflects the strength of social presence in the PeerCollab condition. All participants reported sense of belonging~\cite{ribera2017sense,sense@site}, measured with an adapted 6 item scale --- 1) feeling understood, 2) feeling connected, 3) feeling welcomed 4) being respected by others 5) mattering to others and 6) being happy in the group --- with response levels: never, slightly, somewhat, quite and extremely. The PeerCollab condition saw significantly more sense of belonging, with mean differences significant for each item~\ref{tab:controlstudytab}. The most notable effect comes from items 5 and 6 which reflect PeerCollab community members felt that they mattered to each other, and that they were comfortable in the group setting.

\begin{table}[tb]
 \begin{tabular}{l|ll}
 Scale item & Mean difference & [95\% CI]\\
 \hline
 1. Feeling understood & 1.73 & [1.27, 2.0] \\
 2. Feeling connected & 1.55 & [1.18, 2.0] \\
 3. Feeling welcomed & 1.45 & [0.91, 2.0] \\
 4. Respected by others & 2.09 & [1.64, 2.36] \\
 5. Mattering to others & 2.45 & [2.0, 2.82] \\
 6. Happiness in the group & 2.36 & [1.91, 2.82] \\
 \end{tabular}
 \caption{The unpaired mean differences between the PeerCollab condition and the control condition ($PeerCollab - control$). All p-values in two-sided permutation t-test $\leq 0.01$.}
 \label{tab:controlstudytab}
\end{table}

\subsection{How do people behave with PeerCollab?}
In the field study, PeerCollab was initiated with place holder course spaces for each of 4 MOOC courses, so that participants could build \textit{rapid communities} readily. Participants were to join any course space, or make their own, however at the end of the 6-week study, no new course spaces had been created. Instead participants created 5 independent rapid communities comprising 286 threads in total and consuming 2.4 hours per member on average over the 6 week study. 60\% joined the platform via links posted in the social media such as Facebook and 40\% were direct links by their referrals. 



\subsubsection{Specific behaviours in the communities}
In the \textit{Design thinking} MOOC, 2 communities were created: "Design thinking online community" and "Human Centered design group". Each of the other courses had only one rapid community per course. Although the community guidelines are the same for the entire user base of PeerCollab, rapid communities can have community-level rules and charters developed by the community and headed by the community manager. In this case, the 2 communities in \textit{Design thinking} had different charters reflecting their regular meeting times and discussion agendas. 

Participants could create or reply to public threads, use a "Like" button on posts, and privately any other member from their profile page. The \textit{Creative Problem solving} MOOC community had the most threads, representing 44\% of all discussion during the study. Community 1 in the \textit{Design thinking} had 27\%, and Community 2 there had 15\%. \textit{Introduction to java} had 12\%, and \textit{C++} only received 2\%. Community managers played different roles across communities, contributing 23\% of the conversation on average: in \textit{C++} almost all posts were written by the community manager, while the next highest manager contribution was \textit{Design Thinking} Community 1 at 12\%, then Community 2 at 5\% and lastly the \textit{Java community} manager at 4\%. 

\subsubsection{Social Presence in Communities}
\begin{figure}[tb]
 \centering
 \includegraphics[width=\linewidth]{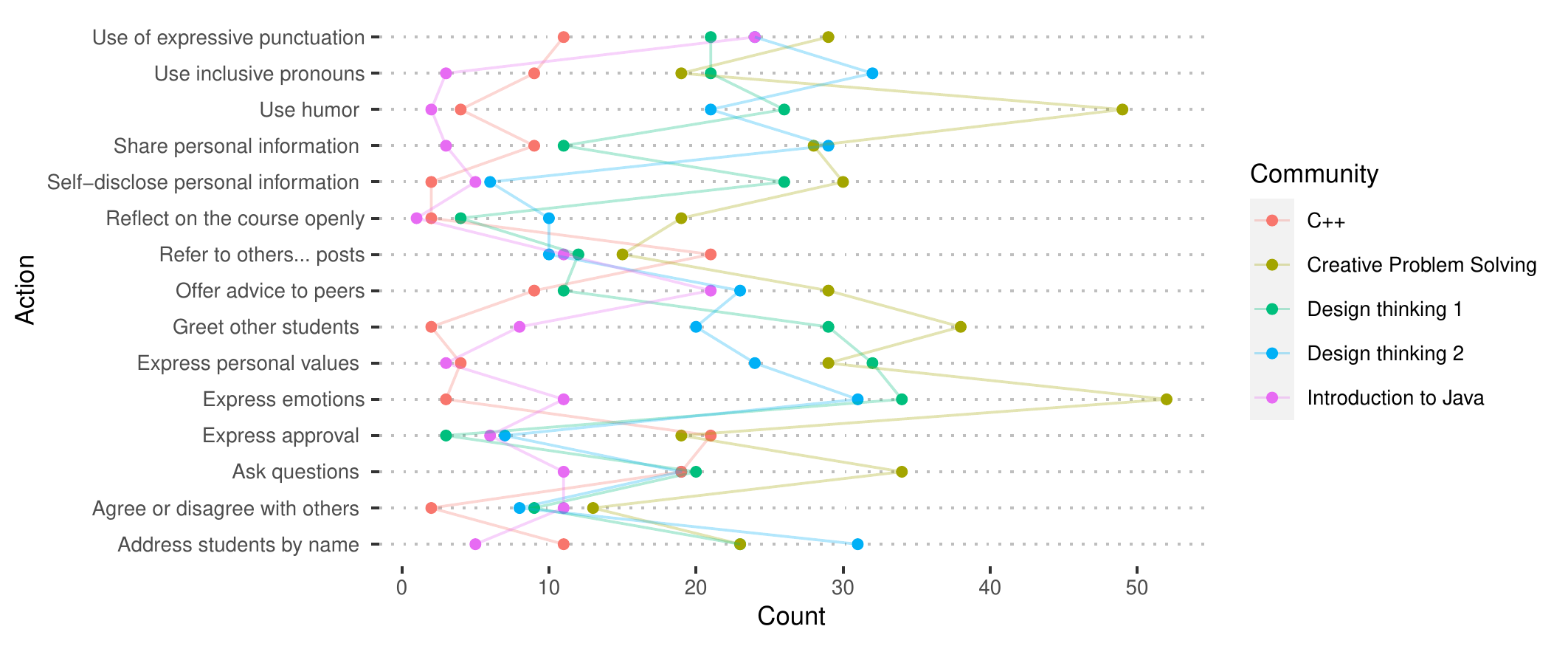}
 \caption{Social presence actions by community}
 \Description{}
 \label{fig:social}
\end{figure}

To understand how these communities used discussion, we analysed engagements (comments and likes) in the threads in each community. We were specifically interested to understand if the communications in the discussions suggested the development of social presence in the communities. Social presence can be identified using a framework from the Communities of Inquiry (CoI) model~\cite{garrison2010exploring} in which community behavior is coded by features identified by several indicators --- for example, \textit{affect indicators} code emotions in the content, \textit{cohesive indicators} code group referencing, using names and sharing, and \textit{interactive indicators} code acknowledgements, question asking and approvals~\cite{wise2004effects}. We coded each communities' data with 15 indicators, and compare the number of occurrences across communities (Fig.~\ref{fig:social}). 


We observed that members across the 5 rapid communities reflected social presence indicators in dramatically different ways. For example, when expressing emotion \textit{Design thinking} Community 1 and 2 had 26\%, 23\% respectively, while the \textit{Creative problem solving} community had 39.7\% emotions and \textit{Introduction to Java} at 8.4\% , \textit{C++} had only 2.3\%. 
Utterances encoded in this way often represent instances of emotion showing the state of members feelings, depicted in the text of their community postings. For example in the \textit{Creative problem solving} community a participant mentioned the following comment which we categorised as a positive state of feeling. 
\begin{displayquote}
~\textit{"....I am glad to see this is happening..."}, 
\end{displayquote}

\subsubsection{Role of Community manager and members and engagements}
Each community in PeerCollab has a community mangers who agrees to lead and facilitate their action. Apart from in the \textit{C++} community, community managers actively engaged with members in each community, with their most frequent actions being regular comments, encouraging comments, and welcoming members. The manager in \textit{Design thinking} Community 1 and \textit{Creative Problem solving} held 2 and 1 online meetings with ZOOM respectively through discussion with their members. These off-platform meetings were unavailable for our analysis because they were outside of the PeerCollab system. In a post interview, the manager of  Community 1 in \textit{Design thinking} reflected~\textit{".. we had 9 participants in the meeting for ~30mints and mainly introduced each other, shared our interests. We could not much discuss course work yet, but we planned our design to share in the forum.."}. Since the role of community manager is likely to be somewhat unfamilar and is not in the normal scope of MOOC participation, we provided explicit expectations and action items to benchmark their activities. During our interviews with the community managers, they reflected the heavy workload they had administrating the work, they particularly highlighted scheduling meetings, and reaching out to members who were not active in the community. 

\begin{displayquote}
\textit{"It was most difficult to navigate who did reply and and who did not, I could not trace since there were many members, I wish the system provide me such details easily than me digging the conversations"} --- Community Manager, Design thinking Community 1
\end{displayquote}

Members behavior reflected appreciation to the community managers. We observed that they were much happy to be addressed personally by name, and welcomed to the community. In post interviews, some members reflected that such addressing lead them to feel accountable to engage and contribute in the discussions. 

\begin{displayquote}
\textit{"I found it cool to be welcomed personally and felt the that this is serious, though sometimes I don't have time to watch the videos in the course I knew I could reply on this space to get summary and insights "} --- Community Member, Creative problem solving community 1
\end{displayquote}

In 3 communities, Design thinking Community 1, 2 and Creative Problem solving, specific behaviours emerged around sharing their work to the community space and receiving comments as feedback. Both these courses had activities which more effective when with feedback, and had creative elements, for example design thinking project prototype ideas. However, in other communities --- C++ and Introduction to java --- very little discussions occurred. Some discussion focused on questions about the code errors, but interestingly, this lead to code snippets being shared, and rarely cultivated other community interaction. This suggests that provided the engagement and facilitation, community members are willing to share artifacts and feedback with confidence. Specifically, the Design thinking Community 1 reflected to phase as a group activity to do activities related to the course. 

\begin{displayquote}
\textit{"Shall we try out to do the assignment and discuss by next week to see who can come up with the model"} --- Community Member, Design thinking community 1
\end{displayquote}

Overall, the field study surfaced insights about how community members were encouraged and led shared activities, how new members were welcomed and, how many participants became confident in sharing their thoughts with the group. However, on the other hand, we also observed that some participants were so occupied watching the videos that they felt the need to apologize to the community for not being more active.

\section{Discussion} 
Through exploring activities in PeerCollab, we learned how rapid communities are created by users, how community managers role-out their assigned functions and how members adhere to the community activities, guidelines and provided structure. We also discovered how rapidly communities come to consensus on activities, negotiate and share learning. Interestingly, we also showed how creating a friendly social environment with active personalized discussions encourages social interactions and collaboration among members, achieving a close-knit feeling, leading to affective, cohesive, and interactive engagements as describe in \textit{communities of inquiry} (CoI). According to \citeauthor{wenger2010communities}, \textit{Communities of Practice} is effective when members of the community adhere to a practice specialised in domain. We created specific spaces as domain by clustering to a limited cohort, as well as a set of interest-aligned members as community and semi-structure through community guidelines, action items, and roles for each community as elements of practice. The membership engagements, facilitator actives and shared artifacts reflect the effectiveness of the community. Although the communities created in this study did not reflect equal levels of engagement, the field study provided us insight on ways to improve the system in the future --- creating more structured instruction for community managers, and creating more community resources for courses that are unlikely to lead to exposure of creative ideas, e.g. the programming courses in our study.

\section{Conclusions}
In this paper, our contributions are twofold. First, our objective was to bring missing social component to MOOCs --- social interactions and sense of community. For this we presented PeerCollab, a socio-technical system designed to build communities for MOOC participants, adding a structured social component while learning. PeerCollab leverages community theories, specifically the communities of practices (CoP) literature. Second, we evaluated the initial instantiating of PeerCollab using a controlled study and field deployment to understand the behavioural effects of this system in MOOCs. Our controlled study resulted in significant levels of social presence and sense of belonging when a PeerCollab was used, instead of the typical MOOC social tools. The field study provided us insights on how different communities reflect their social activities and shared learning together as a loosely cohesive group. While not comprehensive, the examples discussed here illustrate the possible ways to improve effectiveness in MOOC communities for social learning and fill the gaps, supporting MOOCs in bringing a culture of shared active learning, instead of just a culture of isolated content consuming common in MOOCs today. By explicitly designing spaces with semi structured rapid learning communities, we provide the potential of large scale social learning, and enhancing engagement in MOOCs generally. 


\bibliographystyle{ACM-Reference-Format}
\bibliography{reference}


\end{document}